\documentclass[traditabstract]{aa}  

\usepackage{graphicx}
\usepackage{subfigure}
\usepackage{multirow}
\usepackage{enumerate}
\usepackage{color}
\usepackage{natbib}
\usepackage{amsmath}
\usepackage{bbold}
\usepackage{scalefnt}
\usepackage[T1]{fontenc}
\usepackage{mathtools}
\bibpunct{(}{)}{;}{a}{}{,}
\usepackage{caption}
\usepackage{hyperref}
\usepackage{threeparttable}

\begin{document}

\title{Instrumental polarisation at the Nasmyth focus of the E-ELT
}

\author{M.\,de\,Juan\,Ovelar\inst{1}\thanks{email contact: mjovelar@strw.leidenuniv.nl}, F.\,Snik\inst{1}, C.\,U.\,Keller\inst{1} and L.\,Venema\inst{2}}

\institute { 
                 {\inst{1}Leiden Observatory, Leiden University,
                 PO Box 9513, 2300 RA Leiden, The Netherlands}\\                 
                 {\inst{2}ASTRON, P.O. Box 2, 7990AA Dwingeloo, The Netherlands}
                 }

\date{Received date ; accepted date}

\abstract{
The $\sim$39-m European Extremely Large Telescope (E-ELT) will be the largest telescope ever built. This makes it particularly suitable for sensitive polarimetric observations, as polarimetry is a photon-starved technique. However, the telescope mirrors may severely limit the polarimetric accuracy of instruments on the Nasmyth platforms by creating instrumental polarisation and/or modifying {the polarisation} signal of the object. In this paper we characterise the polarisation effects of the two currently considered designs for the E-ELT Nasmyth ports as well as the effect of ageing of the mirrors. By means of the Mueller matrix formalism, we compute the response matrices of each mirror arrangement for a range of zenith angles and wavelengths. We then present two techniques to correct for these effects that require the addition of a modulating device at the ``polarisation-free'' intermediate focus that acts either as a switch or as a part of a two-stage modulator. We find that the values of instrumental polarisation, Stokes transmission reduction and cross-talk vary significantly with wavelength, and with pointing, for the lateral Nasmyth case, often exceeding the accuracy requirements for proposed polarimetric instruments. Realistic ageing effects of the mirrors after perfect calibration of these effects may cause polarimetric errors beyond the requirements. We show that the modulation approach with a polarimetric element located in the intermediate focus {reduces} the instrumental polarisation effects down to tolerable values, or even removes them altogether. The E-ELT {will} be suitable for sensitive and accurate polarimetry, provided frequent calibrations are carried out, or a dedicated polarimetric element is installed at the intermediate focus.
}


\authorrunning{M.\,de\,Juan\,Ovelar et al.}
\titlerunning{Instrumental polarisation at the E-ELT}

\maketitle

\section{Introduction} \label{sec:intro}

The European Extremely Large Telescope (E-ELT) is a $\sim 39$-$\rm{m}$ optical/infrared telescope that {will} take {ground-based astronomy} to the next level {\citep{delabre08,mcpherson12}. Its more than $970\,\rm{m}^2$ of collecting area and unprecedented spatial resolving power will allow for {revolutionary astronomical observations}. Amongst the science goals {of} {the E-ELT} are the study of exoplanets and protoplanetary systems, high redshift galaxies and star formation processes {\citep{hook09,liske12}}. These fields are particularly {demanding}, observationally speaking, and will therefore benefit directly from the technological leap the E-ELT represents. However, photometry, spectrometry and imaging techniques {will not} be able to asses the complete spectrum of open questions without the help of polarimetry {\citep[see, e.g.,][]{strassmeier09}}.

The polarisation state of light retains information about the physical processes by which it is produced (e.g. magnetic fields, reflection and scattering, {inherent asymmetries, etc,} \citealt{tinbergen96, clarke10, snik13}). In addition, polarimetry {boosts} high contrast imaging techniques {by suppressing the flux from the unpolarised central star while keeping the signal from the (polarised) scattering circumstellar matter. This makes it particularly suited for direct imaging and characterisation of exoplanets and the circumstellar discs in which they are born \citep[see][for some theoretical and observational examples]{seager00,stam04,stam08,dekok11,hashimoto11,quanz11,quanz12,quanz13a,dong12,thalmann13,canovas13,dejuanovelar13a}. 
%In the latter case it is particularly crucial to reveal the structure that lies in the closest regions to the host star, where extremely interesting physical processes take place \citep[e.g.][]{jayawardhana06,brogi12b,pinilla12a,dong12,quanz13b,quanz13a,canovas13,dejuanovelar13a}.

%In the former case theoretical simulations such as \citet{seager00,stam04,stam08,dekok11} show that through accurate (spectro-)polarimetry, we can obtain unique information about the structure and composition of their atmospheres and/or surfaces. In the case of discs, polarimetry is already proving its value with observations that reveal deeper and, often previously unseen, details of the circumstellar material \citep{quanz11, hashimoto11,quanz12,canovas12,jeffers12,dong12,thalmann13,quanz13a}. In this case it is particularly crucial to reveal the structure that lies in the closest regions to the host star, where extremely interesting physical processes take place \citep[see][for some examples]{jayawardhana06,brogi12b,pinilla12a,quanz13b}.

Provided a proper instrument design, polarimetry and spectropolarimetry are techniques mainly limited in {sensitivity} \citep[i.e., the noise level for the polarisation measurement,][]{snik13} by the amount of photons collected. However, each element in the optical path can affect the polarisation state of the light coming from the astronomical source limiting the polarimetric {accuracy}  \citep{snik13}. In terms of photon collecting power the E-ELT will be ideal for polarimetry. However, the configuration of the mirrors designed for the Nasmyth focus of this telescope is of particular complexity posing a challenge to perform accurate polarimetry at this location

The {folding} of light to the Nasmyth focus of telescopes is usually achieved by a $90^{\circ}$ reflection on a mirror which generates linear (instrumental) polarisation (IP) signals of a few percent (e.g.\,up to a $5\%$ {at visible wavelengths} \citealt{gehrels60,cox76,joos08,vanharten09,perrin10}). Additionally, a fraction of the incoming linear polarisation {is lost} in the process due to conversion into circular polarisation, which is known as the ``cross-talk'' (CT) between linear and circular Stokes {parameters}. It is known that these instrumental effects can be corrected by further reflection on a second ``twin'' mirror positioned in a ``crossed'' configuration \citep{cox76}. In the case of Nasmyth focus instruments, however, the {mirror} used to deflect the light rotates together with the telescope while the ``crossed twin" usually remains fixed at the Nasmyth port causing this  ``crossed''  configuration to only occur for certain positions of the telescope.

 A retarding element positioned at the entrance of the Nasmyth port can be used to de-rotate the polarisation such that it is always compensated by the ``twin" mirror \citep{sanchezalmeida95,tinbergen07}. This solution has been successfully applied to the design of ZIMPOL \citep[see][]{dejuanovelar12spie}, the polarimeter of the VLT's planet finder SPHERE \citep{gisler04,stuik05,beuzit06,thalmann08,roelfsema10,schmid10}. In the particular case of the E-ELT, this solution is not applicable since the size of the light beam at this location is too large for the currently available {high-quality} retarders. Additionally, in the E-ELT the Nasmyth {folding} is achieved through consecutive reflection on a minimum of two and a maximum of three mirrors instead of one depending on the finally chosen design. %In the particular case that considers three reflections the last mirror is fixed with respect to the other two (see Section\,\ref{sec:configurations} for details).
To perform accurate polarimetry with the E-ELT it is therefore crucial to analyse the polarisation properties of the optical design and either {correct for or calibrate} any instrumental polarisation effects. 

In this paper, we quantitatively characterise the polarisation properties of the two currently proposed Nasmyth optical designs of the E-ELT and analyse two techniques to reduce the instrumental effects. The study is organised as follows. In Section\,\ref{sec:method} we briefly describe the basics of our modeling approach while in Section\,\ref{sec:simulations} we describe the details of the simulations performed. Section\,\ref{sec:results} describes the results obtained and discusses an example of ageing effects on the mirrors after calibration and Section\,\ref{sec:correction} describes the solutions proposed to correct for the instrumental effects found. Finally Section\,\ref{sec:discussion} presents a discussion of the results obtained and the conclusions of our study.}

%%%%%%%%%%%%%%%%%%%%%%%%%%%%%%%%%%%%%%%%%%
\section{Modeling approach} \label{sec:method}
%%%%%%%%%%%%%%%%%%%%%%%%%%%%%%%%%%%%%%%%%%

We use the performance simulator for polarimetric systems code {\sf M\&m's} \citep{dejuanovelar11} to compute the instrumental polarisation effects generated in the optical path of the E-ELT telescope up to the Nasmyth focus. By means of the Mueller matrix formalism, the code calculates the polarisation properties of a given optical system as well as the effects of the measurement process followed. 

{In this} formalism the polarisation state of light is described by  {a $1\times4$ vector} known as the Stokes vector, $\mathbf{S} = (I,Q,U,V)^T$, where $I$ is the intensity, $Q$ and $U$ are linear polarisations in {the $0/90^{\circ}$ and $\pm\,45^{\circ}$ directions} and $V$ is circular polarisation (symbols in boldface denote matrices or vectors). 
%{The convention used in this study to define the orientation of the Stokes vectors will be defined below.}
The effect that an optical element has on the polarisation state of light passing through it, can be described as the product between the incoming Stokes vector ($\mathbf{S_{in}}$) and  a $4\times4$ matrix that accounts for the polarisation properties of the element (i.e.\,a Mueller matrix $\mathbf{M}$),

\begin{equation} 
	\mathbf{S_{out}} = \mathbf{M_{element}} \mathbf{S_{in}}\,.
\label{eq:mueller}	
\end{equation}

The same holds for an optical system composed of several elements,

\begin{equation} 
	\mathbf{S_{\rm{out}}} =\mathbf{M_n}\cdot...\mathbf{M_2}\cdot\mathbf{M_1}\cdot \mathbf{S_{\rm{in}}} = \mathbf{M_{\rm{total}}} \cdot \mathbf{S_{\rm{in}}}\,,
\label{eq:mueller}	
\end{equation}
where $\mathbf{M_n}...\mathbf{M_1}$ represent the Mueller matrices of the $n$ elements of the optical system with $1$ being the first element in the optical path and $n$ being the last.

In order to measure the Stokes components the modulation and demodulation steps need to be included in the process. The first one consists of ``encoding'' the polarisation state of the incoming light in a set of intensity measurements that can be registered by the detector and is usually performed by two elements in the polarimeter: the modulator and the analyser. The former modifies the state of the incoming polarisation, while the latter acts as a polarisation {``filter''}. By changing the position of the modulator in particular steps (i.e.\,{modulation scheme}), one can control which polarisation {($Q$, $U$ or $V$, {or a linear combination thereof})} passes through the analyser and is contained in the measured intensity ($I_{\rm{meas,}i}$, with $i$ ranging from $1$ to $m$ and $m$ being the total number of intensity measurements performed, as well as the {positions/states} of the modulator). This process can be described by the {``modulation matrix''} ($\mathbf{O}$) which then relates the incoming Stokes vector to the $1\times m$ measured intensity vector ($\mathbf{I_{\rm{meas}}}=(I_1,I_2,\dots,I_m)^{\rm{T}}$),

\begin{equation} 
	\mathbf{I_{\rm{meas}}} = \mathbf{O} \mathbf{S_{\rm{in}}}\,,
\label{eq:modulation}	
\end{equation}

where each row in $\mathbf{O}$ is the first row of the $\mathbf{M_{\rm{total}}}$ matrix of the system at each modulation state $m$.
Each component of the {incoming} Stokes vector can then be obtained {from a} linear combination of these $m$ intensity measurements, a process that is known as demodulation,

\begin{equation} 
	\mathbf{S_{\rm{meas}}} = \mathbf{D} \mathbf{I_{\rm{meas}}}\,,
\label{eq:demodulation}	
\end{equation}

which yields the ``measured'' Stokes vector ($\mathbf{S_{\rm{meas}}}$). 

The complete polarimetric {measurement process} (i.e.\,including optical system properties, modulation and demodulation steps) can then be represented by a matrix that is often known as the {``response matrix''} ($\mathbf{X}$, \citealt{ichimoto08}) which relates the incoming Stokes vector with the measured Stokes vector,

\begin{equation} 
	\mathbf{S_{meas}} = \mathbf{X} \mathbf{S_{in}}\,,
\label{eq:measuredstokes}	
\end{equation}

where $\mathbf{X} = \mathbf{D} \mathbf{O}$.

The response matrix is a $4\times 4$ matrix that therefore includes the effect of both the optical system and the defined modulation/demodulation schemes. This makes it a powerful tool for {diagnosing the impact of systematic effects on the polarimetric capabilities of any optical system accounting for the modulation/demodulation processes.

The results obtained in this study are presented in terms of the response matrix and to facilitate their analysis Eq.\,\ref{eq:Xmeaning} shows the relation each of its element represents,

\begin{multline} \label{eq:Xmeaning}
\mathbf{X} =\\
\renewcommand{\arraystretch}{1.5}
\scalefont{0.8}
\begin{pmatrix*}[l]
 I_{in} \to I_{\rm{meas}} 	&\mkern15mu Q_{in} \to I_{\rm{meas}} 	&\mkern15mu U_{in} \to I_{\rm{meas}} 		&\mkern15mu V_{in} \to I_{\rm{meas}}\\
 I_{in} \to Q_{\rm{meas}} 	&\mkern15mu Q_{in} \to Q_{\rm{meas}} 		&\mkern15mu U_{in} \to Q_{\rm{meas}} 		&\mkern15mu V_{in} \to Q_{\rm{meas}}\\
 I_{in} \to U_{\rm{meas}} 	&\mkern15mu Q_{in} \to U_{\rm{meas}} 		&\mkern15mu U_{in} \to U_{\rm{meas}} 		&\mkern15mu V_{in} \to U_{\rm{meas}}\\
 I_{in} \to V_{\rm{meas}} 	&\mkern15mu Q_{in} \to V_{\rm{meas}} 		&\mkern15mu U_{in} \to V_{\rm{meas}} 		&\mkern15mu V_{in} \to V_{\rm{meas}} 
\end{pmatrix*}
\,.
\end{multline}

Diagonal elements {represent the fractional transmission of a} Stokes component throughout the measurement process. Elements in the first column ($I_{in} \to Q,U,V_{\rm{meas}}$) give the polarisation that is generated by the system (IP). Elements relating $Q_{\rm{in,meas}}$ and $U_{\rm{in,meas}}$ are known as rotation while the ones relating $Q_{\rm{in,meas}}$ or $U_{\rm{in,meas}}$ with $V_{\rm{in,meas}}$ give the cross-talk (CT). In presenting our results, we analyse the CT focussing on elements $X_{3,2}, X_{4,2}$ and $X_{4,3} $}. 

%Finally, polarimetric calibration accuracy requirements can be described by as $\Delta \mathbf{X} = \mathbf{X} - \mathbb{1}_4$.

Provided a set of optical elements and the modulation/demodulation schemes, the {\sf M\&m's} code computes all Mueller matrices of the elements and generates the $\mathbf{M_{\rm{total}}}$,  $\mathbf{O}$, $ \mathbf{D}$ and $ \mathbf{X}$ of the system. In obtaining $\mathbf{X}$ the code either computes $\mathbf{D}$ as the inverse or pseudo-inverse of $\mathbf{O}$, depending on the particular case \citep{deltoroiniesta00}, or requires the user to specify it. In the simulations presented in this study we define the demodulation matrix such that it corresponds to an ideal polarimeter. The reason for this is that our aim is to model the behaviour inherent to the optical arrangement and the impact the modulation has on it and not the behaviour of the polarimeter. In this way, the matrix $\mathbf{O}$ includes the realistic behaviour of the elements in the optical system while $\mathbf{D}$ is only computed for the ideal polarimeter. This will cause the response matrix to show the behaviour of the optical system including the modulation scheme but not any effects from the polarimeter.

Some other considerations regarding our simulations are

\begin{enumerate}
	\item{{The dispersion of the index of refraction with wavelength is} included for all materials used. However, for the thin amorphous alumina layer on top of the mirrors a constant value of n=1.6 was assumed, which is an approximation of the value in the studied region ($[500-900]\,\rm{nm}$, \citealt{eriksson81}).}
	\item{{Unless explicitly noted}, all mirrors have the same characteristics, i.e.\,no differential effects are included.}
	\item{Whenever available, real material characteristics and design parameters are {used} to describe optical elements.}
	\item{The Mueller matrices have only been established for the chief ray and therefore the centre of the field of view.} %This is reasonable as polarimetric instruments for the E-ELT will most likely be operating at the center of the field of view.}} %The effects of converging beams are still to be implemented.}
	\item{Only ideal Mueller matrices are taken into account, i.e.\,deviations from the characteristic values of the parameters of optical elements are not included.}
	%meaning that we do not include errors in the individual elements of the optical system.}
	\item{The efficiency of the detector is assumed to be perfect.}% i.e\,the incident and measured {intensities are proportional to each other.}}
	\item{We quantify polarisation effects that can be described with Mueller matrices. We therefore disregard polarisation effects that may be brought about by, e.g., (residual) seeing, differential aberrations, or diffraction effects. As such, the results presented here are for the average of the point spread function (PSF) of the telescope \citep{sanchezalmeida92}}
\end{enumerate}

%%%%%%%%%%%%%%%%%%%%%%%%%%%%%%%%%%%%%%%%%%%%%
\section{E-ELT Nasmyth configurations.}\label{sec:simulations}
%%%%%%%%%%%%%%%%%%%%%%%%%%%%%%%%%%%%%%%%%%%%%
\begin{figure*}[!htb]
	\begin{center}
		\subfigure[]{\label{fig:noM6config}\includegraphics[scale=0.5]{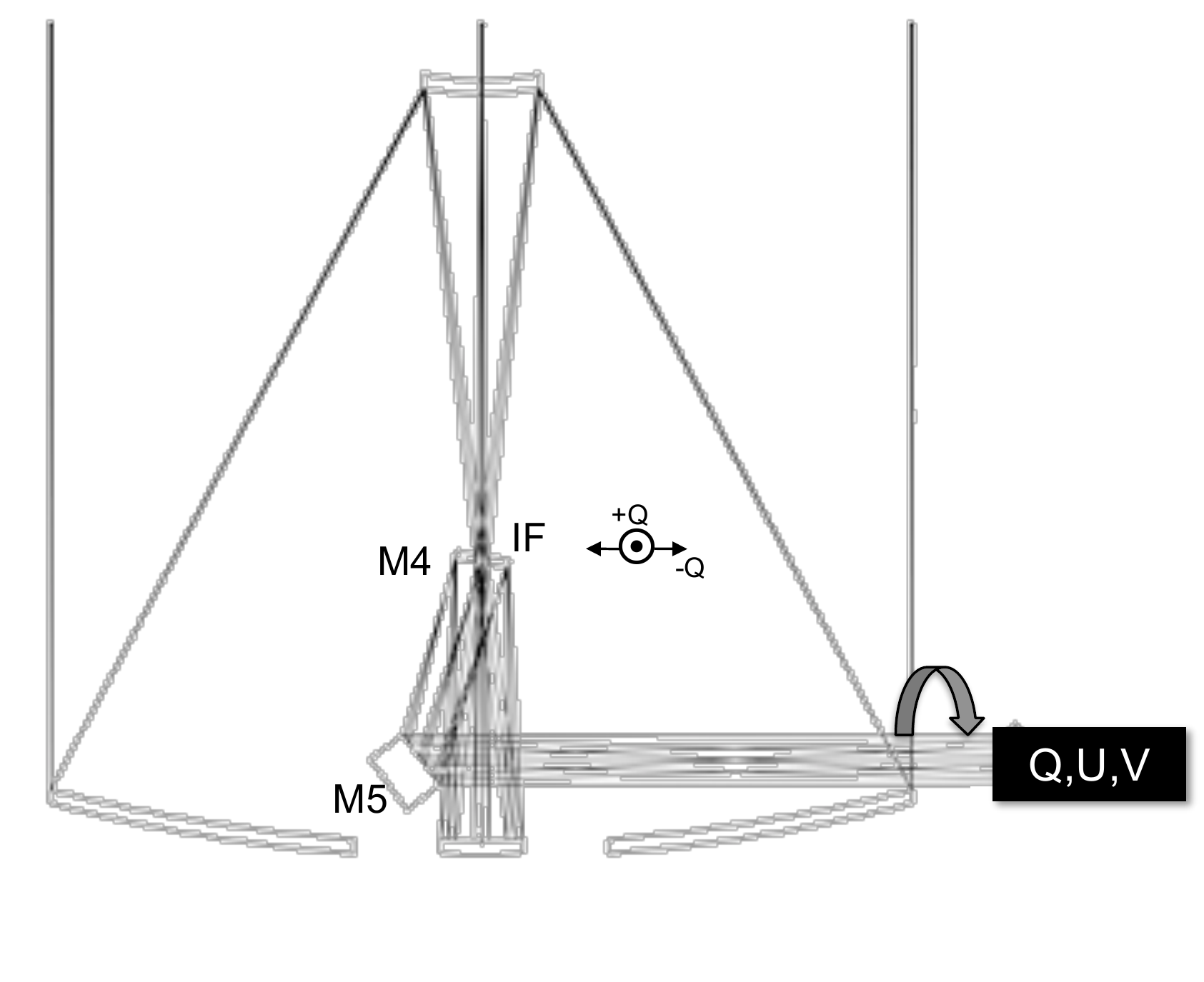}}
		\subfigure[]{\label{fig:fullconfig}\includegraphics[scale=0.5]{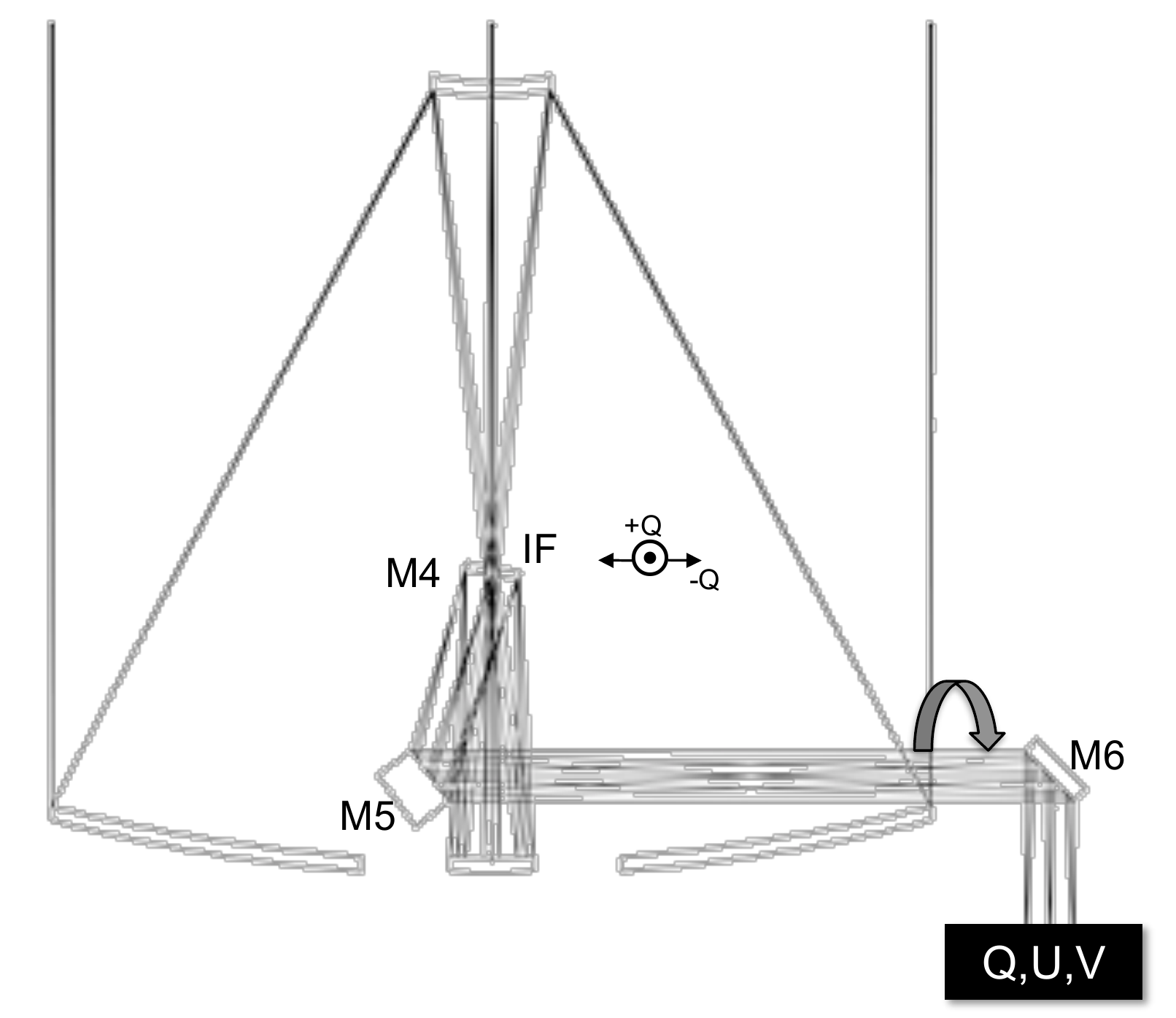}}
	\end{center}
	 \caption{Mirror arrangements considered for the E-ELT Nasmyth configuration with mirrors and intermediate focus (IF) of the telescope marked. (a) {Straight-through} configuration, with the Nasmyth focus directly after the fifth reflection. (b)  {Lateral} configuration, with a sixth mirror fixed in the Nasmyth port. The $Q$, $U$, $V$ box represents a {perfect} full-Stokes polarimeter at the corresponding Nasmyth focus. The rotation axis of the telescope and the $\pm Q$ directions at the intermediate focus position are indicated. Adapted from \citet{delabre08}.
	 } 
	 \label{fig:EELTconfigs}
\end{figure*}

\begin{table}[h]
	% Stretch the rows a bit to give more space
	\renewcommand{\arraystretch}{1.3}
	%\renewcommand{\captionfont}{\scshape}
	% Changes table font (not caption! see above)
	\small
	\begin{center}
		% newline does what you expect
		\caption{Modulation scheme for E-ELT {straight-through} and {{lateral}} configurations\label{tab:modscheme}}
		\begin{threeparttable}
			\begin{tabular}{ccccc}
			\hline
			\hline
			\multicolumn{1}{c}{Modulation} &
			\multicolumn{1}{c} {{\multirow{2}{*}{QWP\tnote{1}}}} & 
			\multicolumn{1}{c} {{\multirow{2}{*}{HWP\tnote{1}}}} &
			\multicolumn{2}{c}{Measured Stokes} \\			
			\multicolumn{1}{c}{state} &
			\multicolumn{1}{c}{$ $} & 
			\multicolumn{1}{c}{$ $} &
			\multicolumn{2}{c}{component} \\
			$m$	& ($^{\circ} $)	& ($^{\circ}) $	& {straight-through}		& {{lateral}} \\
			\hline
			1 	& out 			& 0 				& $\phantom{-}Q$	& $\phantom{-}Q$ \\
			2 	& out 			& 45			& $-Q$				& $-Q$ \\
			3 	& out 			& 22.5			& $-U$				& $-U$ \\
			4 	& out 			& 67.5			& $\phantom{-}U$	& $\phantom{-}U$ \\ 
			5 	& 45  			& 0				& $-V$				& $\phantom{-}V$ \\
			6 	& 45  			& 45			& $\phantom{-}V$	& $-V$ \\
			\hline
			\end{tabular}
			\begin{tablenotes}
			\item[1]{\tiny{QWP and HWP columns show the angle between the fast axis of the wave-plates and the axis of the analyser.}}
			\end{tablenotes}
		\end{threeparttable}
	\end{center}
\end{table}

Figure\,\ref{fig:EELTconfigs} shows the positions of the E-ELT mirrors in the two Nasmyth configurations considered by the current optical design. In the first one (Fig.\,\ref{fig:noM6config}), {straight-through} hereafter, the light is sent to the Nasmyth focus after reflection on the five mirrors fixed to the telescope. These mirrors, therefore, rotate with the zenith angle around the Nasmyth ports as the telescope {tracks}. The second set up considered, {{lateral}} hereafter, adds a sixth mirror fixed at the Nasmyth port (Fig.\,\ref{fig:fullconfig}). The first three mirrors (M1, M2 and M3) {are rotationally} symmetric, which makes their contribution to the instrumental polarisation effects negligible \citep{sanchezalmeida92}. 
%{Moreover, we computed the average polarised point spread function (PSF) of the combination of M1, M2 and M3, including the effect of the segments of M1, and found that even if one of the 798 segments is missing in the outer ring, the IP due to M1 is smaller than $\sim0.003/798 \approx 10^{-6}$.}
%{Even if one of the 798 segments is missing in the outer ring, the IP due to M1 is smaller than $\sim0.01/798 \approx 10^{-5}$}
{Therefore, we only consider the effect of mirrors [M4,M5] or [M4, M5, M6] when simulating the {straight-through} and {lateral} arrangements respectively}. All mirrors are made out of aluminum (index of refraction obtained from \citealt{rakic95}) and have a $4\,\rm{nm}$ Al$_2$O$_3$ layer adopted from the measurements of \citet{vanharten09}. Mirrors M4, M5 and M6 have incidence angles of  $8.5^{\circ}$, $36.5^{\circ}$ and $45^{\circ}$ respectively. We consider a range of telescope zenith angles of $z = [0$ - $90]\,\rm{deg}$ and wavelengths of $\lambda = [500$ - $900]\,\rm{nm}$, and a temperature of $T = 10^{\circ}\textrm{C}$. 
%Table\,\ref{tab:EELTcomp} lists the characteristics and positions of the components used in each of the two simulations.

We define the reference system to be fixed to the telescope which is equivalent to {having} the instrument physically {co-rotating} at the Nasmyth port (e.g.\,pupil tracking), {implementing a half-wave plate {in the} instrument that converts the coordinate system of the telescope to the local one}, or de-rotating the data obtained during the data reduction. The $+Q$ direction is defined as {being} aligned with the $s-$ direction of mirror M4, {see Fig.\,\ref{fig:EELTconfigs}}. {With the $+Q$ direction as a reference, the Stokes $+U$ direction is defined to be rotated clockwise by $45^{\circ}$ as we look into the direction of propagation of the light, {and Stokes $+V$ is defined to be rotating counterclockwise}.} 

The total Mueller matrices of both optical arrangements are then computed by the code as:

\begin{equation} \label{eq:muellernoM6}
\mathbf{M_{\textit{straight-through}} = \mathbf{M_{M5} M_{M4}}}\,,
\end{equation}
and

\begin{equation}\label{eq:muellerfull}
\mathbf{M_{\textit{{lateral}}} = \mathbf{R}(z)\mathbf{M_{M6}}\mathbf{R}(z)\mathbf{M_{M5} M_{M4}}} \,,
\end{equation}
where $\mathbf{M}$ stands for Mueller matrices of mirrors and $\mathbf{R}$ for Mueller matrices of rotations. 
%The rotation of the zenith angle is applied to M6 because the reference system is fixed to the telescope and M6 is fixed to the Nasmyth port. This makes M6 appear as {to be} rotating with respect to the reference system. Note also that the de-rotation before M6 has the same sign as the rotation because there is a change of coordinates performed by the mirror \citep{keller02}.

To simulate the ideal polarimeter we implement a perfect modulator using perfect half-wave and quarter-wave plates (HWP and QWP) to measure Stokes $Q,U$ and $V$, respectively. {The HWP rotates the direction of {the} incoming linear polarisation with respect to {its \textit{fast axis}}. The} QWP transforms {circular polarisation into linear polarisation} depending also on the orientation of its fast axis. The QWP is therefore included in our simulations only for the modulation states where we want to measure $V$. The analyser is a perfect polariser aligned with the $+Q$ direction of the {(rotating)} reference system}.

We then specify a {six}-step modulation scheme to encode the $Q$, $U$ and $V$ Stokes components and an (ideal) demodulation matrix that {recovers} them. While the modulation scheme can be used for both configurations of the E-ELT considered here, the demodulation matrix has to be designed specifically for each case since the additional mirror has an effect on {how the Stokes components are} encoded. Table\,\ref{tab:modscheme} shows the modulation scheme used and the Stokes component that each modulation state encodes for the two setups simulated.

Finally, we consider the following requirements for each element of the response matrix, based on those set for the high-contrast imaging polarimeter E-ELT/EPICS-EPOL \citep{keller10}:

\begin{itemize}
\item linear IP (i.e.\,$I\to Q,U$) $<0.1\%$;
\item transmission of linear polarisation (i.e.\,$Q\to Q$ and $U\to U$)  $>95\%$;
\end{itemize}

and high-resolution spectropolarimeters such as ESPaDOns and HARPSpol \citep{ESPaDOnS,HARPSpol}
\begin{itemize}
\item cross-talk ($Q,U\leftrightarrow V$) $< 1\%$.
\end{itemize}

%%%%%%%%%%%%%%%%%%%%%%%%%%%%%%%%%%%%%%%%%%%%%%%%%%%%%%%%%%%%%%%
\section{Response matrices of the E-ELT Nasmyth configurations and effect of mirror ageing}\label{sec:results}
%%%%%%%%%%%%%%%%%%%%%%%%%%%%%%%%%%%%%%%%%%%%%%%%%%%%%%%%%%%%%%%

\begin{figure*} [!htb]
	\centerline{\includegraphics[scale=0.51]{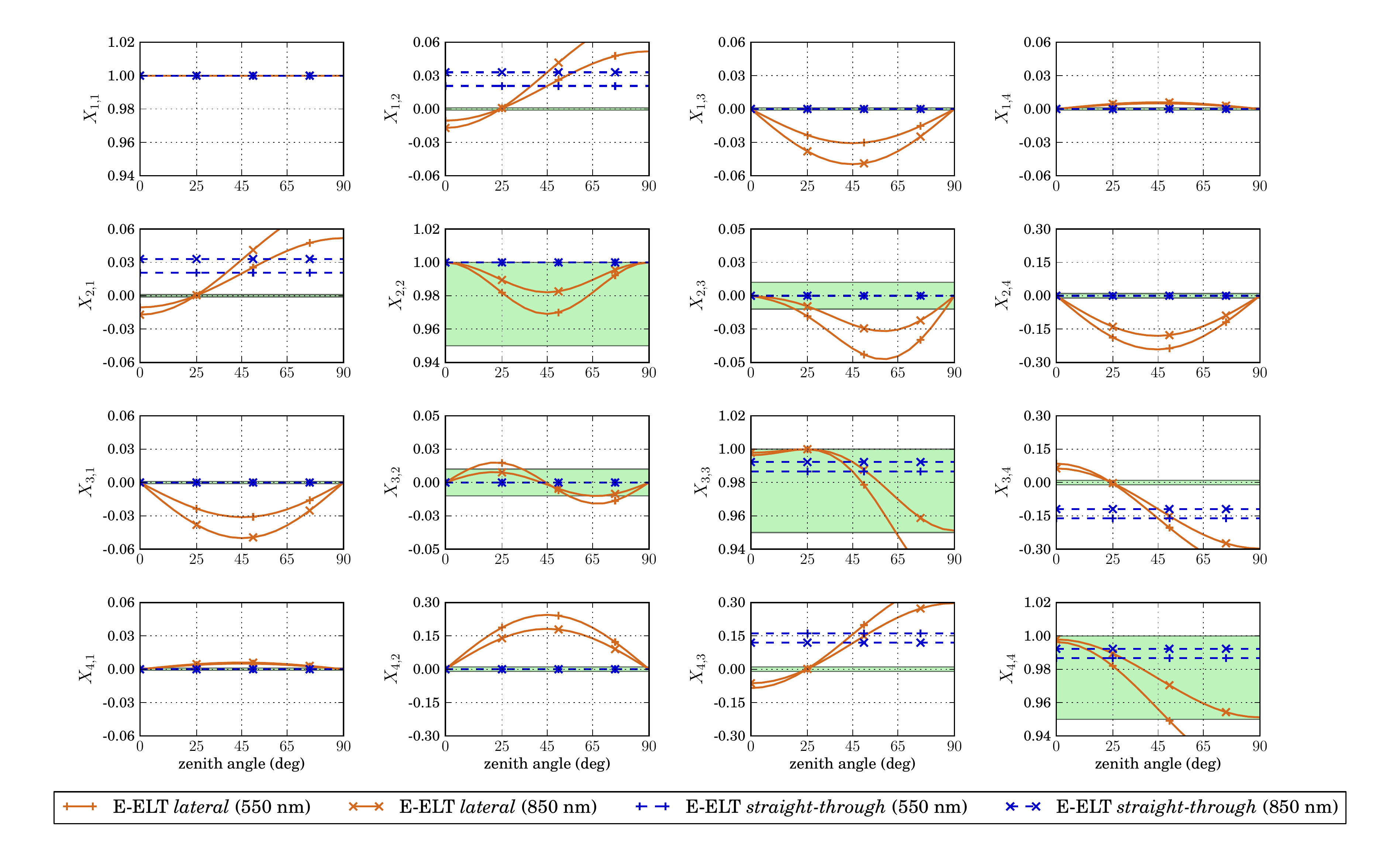}}
	 \caption{Normalised response matrices ($\mathbf{X}$) versus zenith angle ($z$) obtained for the E-ELT's two Nasmyth configurations considered and for the approximated limits of the wavelength range studied ($\lambda = [550, 850]\,\rm{nm}$). Solid and dashed lines correspond to the {lateral} and {straight-through} configurations, respectively. Plus and cross markers denote wavelengths of $\lambda = [550, 850]\,\rm{nm}$, respectively. The light-green {areas} represent the range of values inside the requirements adopted in this study (see Section\,\ref{sec:simulations})}
	 \label{fig:EELT_X}
\end{figure*} 

\subsection{Response matrices}
%==========================

Figure\,\ref{fig:EELT_X} shows the response matrix for the two arrangements studied, normalised to the measured intensity (i.e.\,element $X_{1,1}$), as a function of the zenith angle of the telescope. Solid and dashed lines correspond to the {lateral} and {straight-through} configurations, respectively, while plus and cross markers denote wavelengths of $\lambda = [550, 850]\,\rm{nm}$, the approximated limits of the wavelength range studied. The light-green areas represent the range of values of each element that falls inside the requirements defined above. Note that, in this particular cases, the modulation scheme is such that, when using the ideal demodulation matrix, the response matrix $\mathbf{X}$ coincides with the Mueller matrix of the systems. 

\begin{enumerate}[(a)]
\item{\textit{Straight-through} Nasmyth configuration: \\M4-M5-Nasmyth focus\\
%===============================================================

Dashed blue lines in Fig.\,\ref{fig:EELT_X} show the values of the elements of the response matrix in this configuration. Here, the reference system is fixed to the telescope because mirrors M4 and M5 rotate together with it. This causes the response matrix to be independent of the zenith angle. {This reference system can easily be implemented in any instrument by e.g.\,making the instrument co-rotate with the telescope, placing a retarding element before the instrument capable to de-rotate de polarisation, or de-rotating via the data reduction.}

In terms of instrumental polarisation (elements $X_{1,2}, X_{1,3}, X_{1,4}$) only Stokes $Q$ is generated by this system (element $X_{1,2}$) with values in the range of $\sim [2-3.4]\%$ depending on the wavelength. These values fall well out of the requirements (light-green area).

Because the polarimeter is aligned with the $Q$ direction of the system and it rotates together with the telescope, Stokes $Q$ is transmitted without loss throughout the measurements process (element $X_{2,2}$). The transmission of Stokes $U$ and $V$ ($X_{3,3}$ and $X_{4,4}$) varies depending on the wavelength well within the requirements defined for these elements.

Cross-talk here only {occurs} between linearly polarised light in the $U$ direction and circularly polarised light $V$ ($X_{4,3}$), with values in the range of $\sim [12-16]\%$, outside of the $1\%$ required.}\\

\item{\textit{{Lateral}} configuration: \\M4-M5-M6-Nasmyth focus\\
%===========================================================

Solid yellow lines in Fig.\,\ref{fig:EELT_X} show now the values of the elements of the response matrix in the {lateral} configuration, again with plus and cross markers denoting values for wavelengths of $\lambda = [550, 850]\,\rm{nm}$, respectively. The configuration includes now one more mirror (M6) fixed in the Nasmyth platform. Since the reference system is fixed {with respect} to the telescope (i.e.\,moves together with M4 and M5), the system behaves \textit{``as if"} M6 would be rotating {with} the zenith angle, which introduces a dependency of the response matrix with the zenith angle.

Both linear and circular instrumental polarisation are now generated and vary with the zenith angle. Stokes $Q$ remains outside the specifications for all zenith angles other than $z=25^{\circ}$ (element $X_{1,2}$). Whereas in the case of Stokes $U$ and $V$, the requirements are only met at angles of $z=[0,90]^{\circ}$ (elements $X_{1,3}$ and $X_{1,4}$).

Transmission of all Stokes components (diagonal elements) remains within the requirements except for the case of $U$ and $V$ at long wavelengths and for zenith angles larger than $z>65^{\circ}$ (elements $X_{3,3}, X_{4,4}$).

Here, the cross-talk takes place between both Stokes $Q$ and $U$ and Stokes $V$, and it varies with the zenith angle. In the first case(element $X_{4,2}$) the values fall only inside the requirements for zenith angles of $z=[0,90]^{\circ}$. In the case of cross-talk between $U$ and $V$ ($X_{4,3}$) that only happens for $z=25^{\circ}$.}

\end{enumerate}

\subsection{Effect of ageing of mirrors after calibration} \label{sec:calibration}
%===========================================================
\begin{figure*} [!htb]
	\centerline{\includegraphics[scale=0.5]{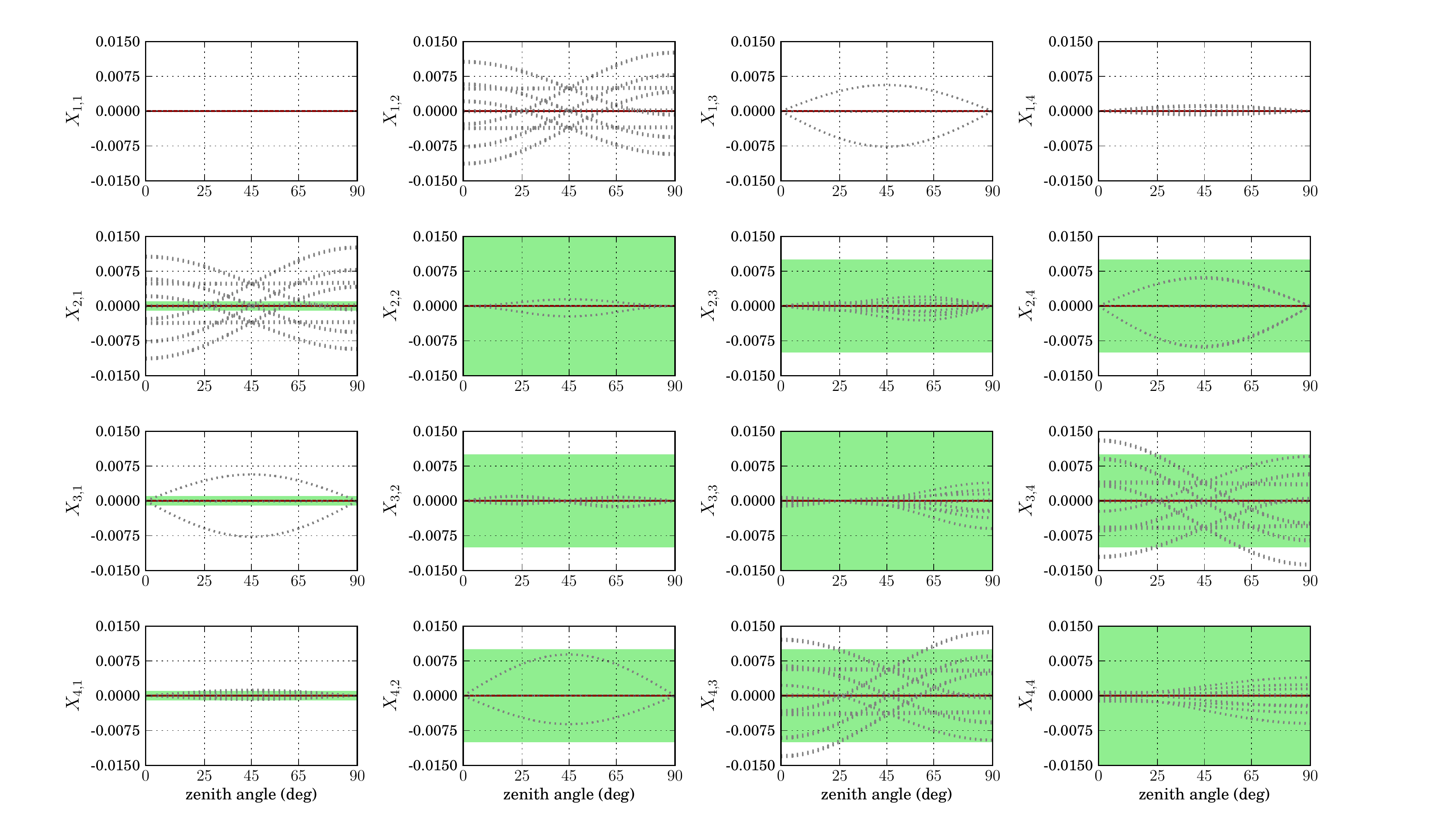}}
	 \caption{Variations in the response matrix of the {lateral} Nasmyth case due to uncalibrated variations in mirror properties. Solid red line shows the values for the elements after calibration at $\lambda= 650$ nm, and dotted grey lines the deviations caused by the ageing. The areas {shaded in green} represent the requirements for polarimetric accuracy (See Section\,\ref{sec:intro}).}\label{fig:comparisonCAL}
\end{figure*}

It is clear that the E-ELT mirrors {produce} instrumental polarisation effects that are in many cases considerably outside of the requirements set. The first question to answer is whether these effects can be calibrated to the required {accuracy}. A comprehensive simulation of calibration procedures is beyond the scope of this paper. {However, for a first estimate of the calibrability, we can compute the impact that variation on mirror properties have on the response matrix. 

For this exercise, we use the {lateral} Nasmyth configuration and assume a perfect calibration of the instrumental effects found. We then vary the properties of M4, M5 and M6 such that it mimics the effects of ageing and mirror pollution, which are the major contributors to the variation in the polarisation properties of telescope mirrors.}To this aim we 1) vary the effective thickness of the dielectric layer on the mirror ($d_f$) by $1\,\rm{nm}$, corresponding to a decrease in mirror reflectivity of $\sim 10\%$ due to the build-up of dust and grime on top of the mirrors, and 2) we vary the absorption term for the mirrors ($\rm{Im(n_m)}$) by 10\%, to represent ageing effects \citep{joos08}. The variation of these two parameters are positively correlated \citep{snik13}. We assume that these variations are independent for M4, M5 and M6, and can go in both directions as mirror cleaning or recoating can take place before or after the calibration.

Fig.\,\ref{fig:comparisonCAL} presents the deviations upon the $\mathbf{X}$ matrix in the {lateral} Nasmyth case for all cases of variations in mirror properties after perfect calibration has taken place. These uncalibrated effects alone make system fall out of the requirements in the case of linear IP and CT between $U$ and $V$. Therefore, frequent calibrations need to take place for polarimetric E-ELT instruments to operate within requirements.

%%%%%%%%%%%%%%%%%%%%%%%%%%%%%%%%%%%%%%%%%%
\section{Correction of the instrumental effects: \textit{switch} and \textit{two-stage modulation} techniques} \label{sec:correction}
%%%%%%%%%%%%%%%%%%%%%%%%%%%%%%%%%%%%%%%%%%

A method for correcting the instrumental effects has to modify the response matrix of the system which, as explained in Section\,\ref{sec:method}, depends on the Mueller matrix of the optical system and on the modulation/demodulation process. The correction, therefore, can be achieved either modifying the instrument, e.g.\,adding elements in the light path that compensate the {polarisation effects} or {adapting} the {polarisation} modulation, or any combination of both. 

The first approach is, mathematically speaking, {equivalent} to intrinsically {modifying} the Mueller matrix of the optical system. {In the second one, additional modulation steps are introduced to separate the instrumental polarisation effects from the source polarisation, and consecutively {be minimised with an additional, differential measurement}.} 

In this section we present two techniques of the second type, the \textit{switch} and \textit{2-stage modulation} techniques. We apply them to the correction of the instrumental effects generated on the E-ELT {lateral} configuration, found to be the least optimal for performing accurate polarimetry. In both cases, retarding elements are placed in the intermediate focus of the telescope (see below) to modify the modulation scheme. To fully characterise the effect of these elements in the measurement process, they are always simulated as realistic wave-plates, while the retarders used for the polarimeter are still simulated as ideal elements. The $\mathbf{D}$ matrix is still computed for an ideal polarimeter.

\subsection{The \textit{switch} technique}\label{subsec:switch}
%=================================================

\begin{table}
	% Stretch the rows a bit to give more space
	\renewcommand{\arraystretch}{1.3}
	%\renewcommand{\captionfont}{\scshape}
	% Changes table font (not caption! see above)
	\small
	\begin{center}
		% newline does what you expect
		\caption{Modulation scheme for E-ELT {{lateral}} configuration (M4-M5-M6) + {IF-switch} \label{tab:ifs1scheme}}
		\begin{threeparttable}
			\begin{tabular}{ccccc}
			\hline
			\hline
			Modulation 	& \multirow{2}{*}{{HWPif}\tnote{1}} 	& \multirow{2}{*}{{QWP}\tnote{1}} 	& \multirow{2}{*}{HWP\tnote{1}} 	& Measured \\ 
			state  		&										&										&								& Stokes\\		
			m			& ($^{\circ} $)	& ($^{\circ} $)		& ($^{\circ}) $	& component \\
			\hline
			1 			& 0  			& out 			& 0 			&\phantom{-}Q \\
			2 			& 0  			& out 			& 45			& -Q \\
			3 			& 0  			& out 			& 22.5		& -U \\
			4 			& 0  			& out 			& 67.5		&\phantom{-}U \\ 
			5 			& 0  			& 45  			& 0			& -V \\
			6 			& 0  			& 45  			& 45			&\phantom{-}V \\
			7 			& 45  		& out 			& 0 			&-Q \\
			8 			& 45 			& out 			& 45			&\phantom{-}Q \\
			9 			& 45 			& out 			& 22.5		&\phantom{-}U \\
			10 			& 45 			& out 			& 67.5		& -U \\ 
			11 			& 45 			& 45  			& 0			& -V \\
			12 			& 45 			& 45  			& 45			&\phantom{-}V \\
			\hline
			\end{tabular}
			\begin{tablenotes}
			\item[1]{\tiny{, {HWPif}, QWP and HWP columns show the angle between the fast axis of the wave-plates {and the defined $+Q$ axis}.}}
			\end{tablenotes}
		\end{threeparttable}
	\end{center}
\end{table}

{The \textit{switch} technique is a simple modulation-related technique to apply \citep{tinbergen96,stuik05}. The idea is to implement a rotatable wave-plate as early in the light path as possible to ``switch" the sign of the incoming polarisation while keeping the instrumental effects, generated downstream, fixed. In the case of the E-ELT, this element could be installed in the intermediate focus (IF) of the telescope (see Figure\,\ref{fig:EELTconfigs}). Light passes through this IF on its way from M2 to M3 and therefore, before reaching M4, which makes this focus ``polarisation-free". {This wave-plate, the {switch} hereafter, rotates the direction of the either linear or circular polarisation coming from the sky and the telescope up to this point (i.e.\,sky-M1-M2-M3), alternatively between two orthogonal positions, thus changing its sign. However, the polarisation generated along the optical path of the telescope below (i.e.\,M4-M5-M6) remains unrotated. By taking two sets of measurements with the {switch} in these two positions and subtracting them, one can ideally suppress most of the instrumental polarisation generated in the $Q$ and $U$ directions. 

A potential disadvantage of this technique comes from the fact that the two measurements are taken with a delay in time. If the measurements are separated in time they might end up being slightly different and the subtraction is not {perfect} anymore. Therefore, the technique benefits from {a rapid switching/modulation duty cycle}. 
%This technique can be applied for correcting linear instrumental polarisation in $Q$ and $U$, switching the HWP between {$0^{\circ}$ and $45^{\circ}$ modulo $90^\circ$} (with respect to the analyser axis), respectively. 
%\textcolor{red}{Since making these four sets of measurements is more time-consuming, it is a matter of the particular instrument and design priorities to decide how it is more convenient to use the switch.
%Explain...}   

To show how this arrangement would correct the linear IP in the {lateral} case, we simulate the {IFswitch} with a HWP at the intermediate focus (HWPif) rotating between $0/{45}^{\circ}$ and therefore correcting the instrumental polarisation generated in the $Q$ direction. Table\,\ref{tab:ifs1scheme} shows the resulting modulation scheme. This element is simulated using realistic specifications of an achromatic HWP. The element is composed of two crossed birefringent plates made of quartz and magnesium fluoride (MgF$_{2}$) with thicknesses $t_{quartz}=841.2\,\mu\rm{m}$ and $t_{MgF_2}=674.8\,\mu\rm{m}$. These two plates together {comprise} an achromatic HWP with a working range of $\lambda=500-900\,\rm{nm}$ {centred} at $\lambda=650\,\rm{nm}$. Refractive indices for quartz and magnesium fluoride were obtained from \citet{ghosh99} and \citet{bass09}, respectively.

The wavelength range of the HWP is the limiting factor of this solution since, {any deviation from a perfect half-wave plate will affect the switching performance}. Table\,\ref{tab:ifs1scheme}, shows the positions of all elements involved in the modulation.

\subsection{The \textit{2-stage modulation} technique}\label{subsec:2sm}
%========================================================

\begin{table}
	% Stretch the rows a bit to give more space
	\renewcommand{\arraystretch}{1.3}
	%\renewcommand{\captionfont}{\scshape}
	% Changes table font (not caption! see above)
	\small
	\begin{center}
		% newline does what you expect
		\caption{Modulation scheme for E-ELT {{lateral}} configuration (M4-M5-M6) + {2-stage modulation} \label{tab:2smscheme}}
		\begin{threeparttable}
			\begin{tabular}{cccccc}
			\hline
			\hline
			Modulation 	& \multirow{2}{*}{{HWPif}\tnote{1}} 	& \multirow{2}{*}{{QWPif}\tnote{1}} 	& \multirow{2}{*}{HWP\tnote{1}} 	& Measured \\ 
			state  		&										&								&								& Stokes\\		
			m			& ($^{\circ} $)	& ($^{\circ} $)		& ($^{\circ}) $	& component \\
			\hline
			1 			& 0  			& out 			& 0 			&\phantom{-}Q \\
			2 			& 0  			& out 			& 45			& -Q \\
			3 			& 22.5  		& out 			& 0			& \phantom{-}U \\
			4 			& 22.5  		& out 			& 45			&-U \\ 
			5 			& 45  		& out			  	& 0			& -Q \\
			6 			& 45 			& out			 	& 45			&\phantom{-}Q \\
			7 			& 67.5  		& out 			& 0 			&-U \\
			8 			& 67.5 		& out 			& 45			&\phantom{-}U\\
			9 			& out 		& \phantom{-}45 	& 0			&\phantom{-}V \\
			10 			& out 		& \phantom{-}45 	& 45			& -V \\ 
			11 			& out 		& -45  			& 0			& -V \\
			12 			& out 		& -45  			& 45			&\phantom{-}V \\
			\hline
			\end{tabular}
			\begin{tablenotes}
			\item[1]{\tiny{{HWPif}, {QWPif} and HWP columns show the angle between the fast axis of the wave-plates {and the defined $+Q$ axis}.}}
			\end{tablenotes}
		\end{threeparttable}
	\end{center}
\end{table}

The \textit{switch} technique solves the issue of the polarisation generated in the system for either linear or circular polarisation (not both). However, the same principle can be taken further to develop a \textit{2-stage modulation} technique. In this case, {a full-blown polarisation modulator is located in a ``polarisation-free'' location upstream (in this case, again, the IF of the E-ELT). This modulator converts the measureable polarisation into a polarisation state that is mostly or fully transmitted by the optical system behind it, known as the ``eigen-vector'' of the system \citep{lopezariste11, snik13}. For relatively simple cases, this eigen-vector is linear polarisation, e.g.\, $+Q$ for the straight-through Nasmyth port case. In general, this eigen-vector is elliptical, and varies with the instrument configuration, e.g.\,the pointing in the {lateral} Nasmyth case.

We apply such a two-stage modulator for the E-ELTe placing a modulator in the IF that consists of two achromatic wave plates: HWPif and QWPif. The QWPif is implemented in the same way as the HWPif of subsection\,\ref{subsec:switch} by modifying the thicknesses of the two layers, $t_{quartz}=421.1\,\mu\rm{m}$ and $t_{MgF_2}=337.8\,\mu\rm{m}$, to make it a quarter-wave plate.  These plates (HWPif and QWPif) are used to modulate $Q$, $U$ and $V$ in {a classical way by sequentially converting those polarisation states into $Q$}. This direction is not an eigen-vector for the {lateral} Nasmyth case, but it comes sufficiently close. The polarimeter on the Nasmyth platform now only measures Stokes $Q$, which allows for a much faster duty cycle than the first modulator in the IF. Table\,\ref{tab:2smscheme} shows the position of the elements for this modulation scheme. 
}

\subsection{Corrected response matrix}\label{subsec:2sm}
%========================================================
\begin{figure*} [!htb]
	\centerline{\includegraphics[scale=0.51]{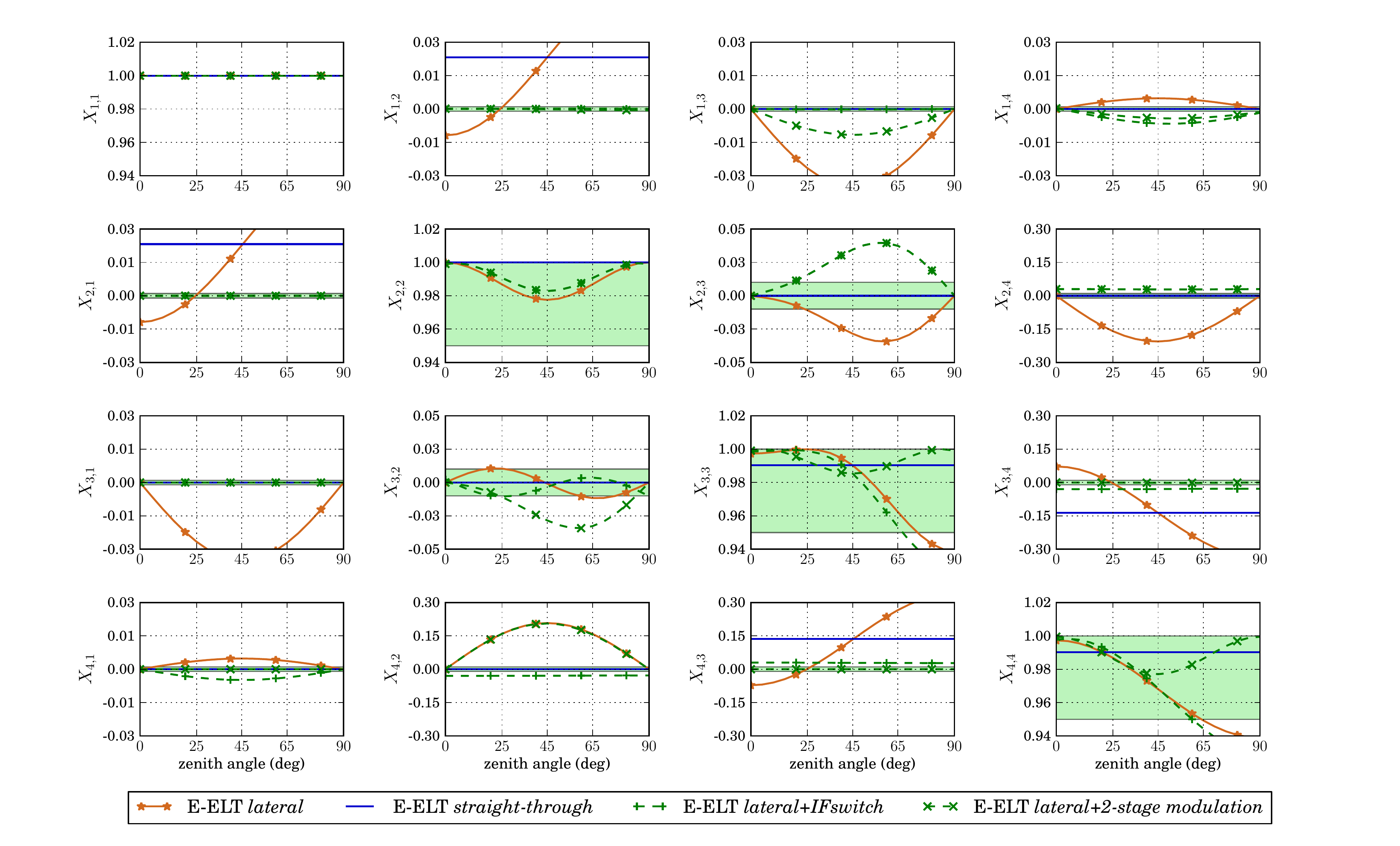}}
	 \caption{Normalised response matrices ($\mathbf{X}$) versus zenith angle ($z$) obtained for the four configurations simulated at $\lambda=650\,\rm{nm}$. Solid, dotted-dashed, dotted and dasehd lines correspond to the {{lateral}}, {straight-through},  {straight-through+IFswitch} and {{lateral}+IFswitch} configurations.\label{fig:comparison}}
\end{figure*} 

Figure\,\ref{fig:comparison} shows a comparison of {the four configurations analysed in this study,} {i.e.\,E-ELT {{lateral}}, E-ELT {straight-through}, E-ELT {{lateral}+IFswitch} and E-ELT {{lateral}+2-stage modulation}. The elements of the four response matrices are shown for a fixed wavelength of $\lambda=650\,\rm{nm}$, which is the centre of the wavelength range studied. Solid lines correspond to the original {straight-through} and {lateral} matrices (blue unmarked and yellow star-marked, respectively) while dashed green lines correspond to the corrected ones, with plus and cross markers denoting values for {{lateral}+IFswitch} and {{lateral}+2-stage modulation}, respectively.

Linear IP is completely suppressed by the two methods (elements $X_{1,2}$, and $X_{1,3}$) while circular IP is only compensated for in the {{lateral}+2-stage modulation} case. The CT between $Q$ and $V$ is improved only in the {{lateral}+IFswitch} solution while that taking place between $U$ and $V$ is significantly improved in this same case and completely corrected for in the {{lateral}+2-stage modulation} arrangement. This technique also improves considerably the transmission of all Stokes vectors while the {{lateral}+IFswitch} only meet the requirements for zenith angles of $z\le65$. The rotation element, $X_{3,2}$, presents, however, a much worse behaviour for the {{lateral}+2-stage modulation} case.

%%%%%%%%%%%%%%%%%%%%%%%%%%%%%%%%%%%%%%%%%%
\section{Discussion and conclusions} \label{sec:discussion}
%%%%%%%%%%%%%%%%%%%%%%%%%%%%%%%%%%%%%%%%%%

The results shown in Fig.\,\ref{fig:EELT_X} show that, according to the requirements set, none of the two Nasmyth configurations being considered for the E-ELT are {suitable} for performing accurate polarimetry. However, the instrumental effects generated in the {straight-through} configuration, i.e.\,IP in the $Q$ direction and CT between $U$ and $V$, are at least independent of the {pointing} of the telescope {when adopting a coordinate system that co-rotates with the telescope} which makes it relatively easy to correct for (e.g., using a tilted glass plate to compensate the IP).

%According to the requirements values, none of the two configurations are {suitable} for performing accurate polarimetry, although it is clear that the \textit{straight-through} arrangement is considerably easier to correct for. Indeed, {when adopting a co-rotating coordinate system for an instrument on the {straigh-through} Nasmyth platform, the instrumental polarisation is constant, and can be corrected for with, e.g., a tilted glass plate (at least for a certain wavelength range).}

Depending on the science requirements, calibration can be a good enough solution to the problem. However, the time required to perform calibration measurements for polarimetry is considerably long due to the fact that photometric and polarimetric measurements are needed as well as a similar signal to noise ratio in both the science and calibration observations. As an example, in current polarimetric observations of circumstellar environments, up to a $50\%$ of the observing time can be lost between calibration with polarimetric standard stars and overheads.

The frequency with which calibration measurements have to be performed is also an important parameter to account for since, as shown in Section\,\ref{sec:calibration}, small variations in parameters of the optical elements can quickly impact the quality of the measurements. 

%{Furthermore, at this point there is no way to guarantee that calibration on standard stars will fundamentally be sufficient to reduce the instrumental polarisation issues due to the E-ELT mirrors to within the values of the error budget for the complete instrument.}
%{This error budgeting for the E-ELT/EPICS-EPOL instrument will be explored in a forthcoming paper.}
%Daytime calibration techniques \citep{harrington11} are a good solution although it is unlikely that such techniques could be applied to E-ELT instruments due to the high risk that opening the dome during day time implies for the telescope. 
%In the E-ELT case, calibration is therefore not enough to deliver the quality of polarimetric measurements required unless one is willing to sacrifice considerable amounts of observing time.

It is in the light of this conclusion that we propose alternative solutions based on extended modulation approaches to correct for the instrumental effects. The {switch} and {two-stage} modulation techniques are applied to the \textit{lateral} Nasmyth configuration and their response matrices computed for the same range of zenith angles and wavelengths considered before. Figure\,\ref{fig:comparison} shows how most of the instrumental effects are stabilised and/or corrected for. In general, the {switch} technique works very well for systems that are focussed on the measurement of linear polarisation while the {2-stage modulation} has the potential of taking care of circular polarisation issues as well. {An additional advantage of the latter implementation is the better response of the system to the measurement of $U$ (see element $X_{3,3}$ of Fig.\,\ref{fig:comparison}) In short, the {2-stage modulation} improves the efficiency of the polarimeter by tuning the eigen-vector of the system to the Stokes component that is being measured.} This, in the framework of our simulations, comes at the price of increasing the rotation between $Q$ and $U$ and the CT between $Q$ and $V$. However, it is important to remark that our modulation scheme is just an example and that different modulation schemes can be optimised to compensate for the particular instrumental effects a given observation {has} to deal with. 

These effective and versatile techniques require the addition of retarding elements to the optical path of the telescope which, considering the already complex optical design of the E-ELT, may be a disadvantage. However they also have the advantage of decreasing the calibration time required. {The retarding elements used in the cases presented in this study also introduce a higher dependency of the values of the response matrix with wavelength, although this depends strongly on the design of the retarder/s and, in principle, it is feasible to taylor them to suit the requirements of a particular case.}

In summary, the E-ELT poses considerable challenges to performing accurate polarimetry, but with the current state of polarimetric techniques it is definitely possible to achieve this goal.

\begin{acknowledgements}
The authors are grateful to Tim van Werkhoven, Visa {Korkiakoski}, David Harrington and Gerard van Harten for insightful discussions and to the anonymous referee for a very useful report that helped improving this study.
\end{acknowledgements}

%%%%% References %%%%%
\bibliographystyle{aa}
\bibliography{/Users/mj/Documents/Bib/mjsbib.bib}

\end{document}